# Concurrent Crossmodal Feedback Assists Target-searching: Displaying Distance Information Through Visual, Auditory and Haptic Modalities

Feng Feng · Tony Stockman

**Abstract** Human's sense of distance depends on the integration of multisensory cues. The incoming visual luminance, auditory pitch and tactile vibration could all contribute to the ability of distance judgement. This ability can be enhanced if the multimodal cues are associated in a congruent manner, a phenomenon has been referred to as crossmodal correspondences. In the context of multisensory interaction, whether and how such correspondences influence information processing with continuous motor engagement, particularly for targetsearching activities, has rarely been investigated. This paper presents an experimental user study to address this question. We built a target-searching application based-on a Table-top, displayed the unimodal and crossmodal distance cues concurrently responding to people's searching movement, measured task performance through usability evaluation, and analysed movement feature through kinematic evaluation. We find that the crossmodal display and audio display lead to improved searching efficiency and accuracy. More interestingly, this improvement is confirmed by kinematic analysis, which also unveiled the underlying movement features that could account for this improvement. We discussed how these findings could shed lights on the design of assistive technology and of other multisensory interaction.

**Keywords** Crossmodal display; Crossmodal correspondences; Target-searching; Kinematic features

## 1 Introduction

In assistive technology designed for map exploration or navigation, audio and/or haptic displays have been used as alternative ways to present distance information [12,40,7]. In these display approaches, the auditory dimensions of amplitude, pitch as well as tactile intensity are frequently used to represent distance [16, 6,7]. However, in most everyday situations, we perceive distance as an integrated product of multisensory cues [57], with either congruent or incongruent combinations between, for example, visual contrast, loudness and/or vibration intensity [22,54]. In this regard, how people combine and react to integrated distance cues with continuous motor skills in navigation-related tasks remains uncharted.

Crossmodal correspondence is a perceptual feature which refers to the associated perceptual relationship between two or more sensory modalities. For example, where a dark object feels heavier than a bright object, and an increase or decrease in pitch is associated respectively with a rise or fall in a vertical position. These crossmodal associations have long been investigated through behavioural and cognitive neuroscience studies. It is well-acknowledged that such associations modulate perception, recognition, and processing with respect to our physical environment [22,17,11]. Many of the congruent correlations are derived from everyday situations when we interact with the physical world [56, 29,43]. For example, there exists natural environmental statistics that reveal a clear mapping between pitch and elevation, and potentially between pitch and brightness, as well as between many other physical properties [46, 36,1,43].

These crossmodal associations have provided fertile ground for the design of crossmodal information displays [30,39,2], including accessible technology [35, 39], embodied learning systems [5], user experience and interaction engagement studies [52,41]. However, the majority of the behavioural research on crossmodal correspondences has been mainly based on the experimental paradigm of speeded classification (and discrimination, identification as well), which is designed to understand human's information processing by providing either congruent or incongruent crossmodal stimuli, and ask people to make a instant judgement on the signal's physical value. These values were chosen to be at the opposite ends of the value spectrum, for instance, high and low for auditory pitch, bright and dark for visual brightness [22,54,18]. With a focus on information processing efficiency, this approach, in general, requires judgements that do not involve intense motor activity.

The focus of the present paper is activities involving continuous motor engagement, particularly searching behaviour performed with the arm. In this regard, the discrete display on polarised values, as mentioned above, have limited applicability in more ecologicallybased interaction scenarios, particularly in ones involving intense motor activity such as percussion performance [15] or navigation-related tasks [6,16]. Tangible interfaces, in a sense, have innate physical dimensions, such as position, colour and acoustic features, that can be leveraged to updating feedback in response to motor activities [31,5,8]. By displaying the distance information based on crossmodal correspondences, we can observe the potential influence that the crossmodal effect has on virtual navigation-related tasks, in our particular case, the target-searching task on a tangible interface.

The research questions addressed in this paper are concerned with exploring whether and how the concurrent and continuous crossmodal display of distance information in a tangible interface can influence people's target-searching performance. In this paper, we present an interactive Table-top instantiation for the changing distance between the current location and the target. We employ usability and kinematic evaluation methods to answer the 'whether' and 'how' questions respectively. Following this analysis, we discuss the relevance of the research in terms of designing assistive technology and for other multimodal interaction scenarios.

The novelty and contributions of the current research are first, to examined the effect of crossmodal correspondences based on continuous changes of the stimuli; second, to explored the influence of concurrent crossmodal feedback on a target-searching task, which required a continuous motor engagement; third, applying kinematic analysis as a way to decode and understand the relationship between target-searching efficiency and movement features.

**2 Background**

2.1 Crossmodal Integration and Crossmodal Correspondence

When we 'pick up'information or perceive affordance from the surrounding environment, our brain processes multisensory information not in a separate and distinct way, but in an affected and integrated manner [32,37, 61]. One of the most investigated perceptual phenomena resulting from this integration process is crossmodal correspondence [54].

For many years, the nature of this phenomenon and its potential benefits on people's information processing ability has been investigated in both the cognitive science research field and Human-Computer Interaction (HCI) scenarios [13,38]. It has been shown that there are different factors that account for this perceptual correspondence. Early cognitive neuroscience studies have shown that spatial and temporal coincidence of stimuli onset modulates crossmodal correspondence [53, 21,58]. Other than spatio-temporal factors, accumulated behavioural studies have shown that semantic and synaesthetic congruency in multisensory perception also influences crossmodal information processing. Semantic congruency refers to a situation in which 'pairs of visual and auditory stimuli are presented that are varied in terms of their identity and/or meaning ', while synaesthetic congruency refers to 'correspondences between more basic stimulus features 'such as pitch corresponding with brightness [54].

A growing number of research studies have investigated plenty of crossmodal correspondences that are synaesthetically congruent, such as auditory pitch is correspondent with both visual brightness, visual spatiality, and visual size [22,50,25]. Congruent synaesthetic correspondences can facilitate people's perceptual performance and cognitive performance such as information classification, discrimination, and working memory capacity [50,22,11].

In spite of the perceptual and cognitive benefits of crossmodal congruency that have been observed under conventional experimental paradigms, i.e. the speeded classification paradigm and unspeeded psychophysical studies, two experimental constraints limit the applicability of the results' to virtual navigation-related tasks. Firstly, the crossmodal information presented in the
speeded paradigm were explicitly or arbitrarily introduced to participants rather than being invoked by their active input behaviour; secondly, the crossmodal information was displayed as a one-time stimulus rather than as continuous feedback in response to participants' motor behaviour.

A recent mobile game application tested the crossmodal congruency effect based on sequences of unimodal or crossmodal stimuli with consecutive input activities. In the game, a sequence of visual-auditory stimuli are first displayed on the screen once, users were expected to reproduce the sequence by tapping on the screen in the right order. Results of user's motor performance revealed that the congruency effect could not be found in the crossmodal condition, possibly due to the compensation of better auditory sensibility [41]. The current study extends this line of research by providing continuous crossmodal display in response to people's step-by-step input behaviour, with a focus on targetsearching tasks that could facilitate visually impaired people explore unknown table-tops.

2.2 The Ecological Approach To Multimodal Perception

While studies involving multisensory perception are usually concerned with single stimuli [20], correlation detection [42] or temporal-spatial multisensory binding [14,55], other research studies provide different perspectives on sensory perception based on activity theory. The ecological view of perception puts the research focus on what affordances can people perceive, rather than optical properties [28]. Inspired by this ecological perspective, Gaver proposed a new framework for audio perception studies, which describes the sound in terms of audible source attributes rather than physical dimensions [26]. Gaver stated that the sounds we perceive are not the sound of the object itself, but what we hear is information about material interactions at a location in an environment. A sound of an approaching engine on the road provides information about the approximate distance of an approaching car, the smoothness of operation of its engine and some qualities of the surface along which its tyres are moving (dry, wet, on gravel).

The ecological view of perception makes crossmodal display a promising approach in the design of assistive technology for the requirement of sensory substitution [39,16], perceptual enhancement [59,23], and more efficient data exploration [24,45]. This perspective requires the consideration of congruent multisensory stimuli, which do not only occur at the level of 'inter-modality' but also at the level of embodied experience [34], especially in the situation when multisensory stimuli are congruent with our previous activities, to which some of the congruent crossmodal correspondences can be traced back [46,33]. This ecological perspective also echoes the statistical correlation of crossmodal perception that existed in the natural environment, and that dynamically update with interaction experience [1,56,43]. On the application level, for example, increased pitch values were used as an indicator of approaching to victims in the abstract version of rescuing operation tasks [45]. In another case, the experiential basis of modalities has been taken into account for formulating design principles for data-exploration tasks [24]. Despite the practical values of the multimodal display, the mutual effect between modalities in target-searching tasks, especially the crossmodal correspondences have rarely been investigated.

3 Investigation Rationale and Design Implementation

Our hypothesis is that the embodied experience of crossmodal correspondence can be spontaneously triggered by leveraging the physical features on the tangible interface, and that it can influence people's target-searching behaviour when displaying the distance concurrently. Thus we propose to use the approach of crossmodal display to render the distance information in a tangible interactive setting. We wished to address the questions of whether and how crossmodal display of distance information can have an influence on target-searching behaviours, while, at the same time, investigating which display strategy better facilitates distance perception and interaction performance. The experiment involved comparisons of strategies employing single-mode displays, single vs. crossmodal displays and comparisons between different combinations of crossmodal displays. This research was given ethical approval by (Blind for review), approval number (Blind for review).

To provide dynamic feedback, systematic gradation of congruent crossmodal stimuli is an important factor to ensure a comparable level of perceptual sensitivity across multi-modalities. Previous research used the method of distance measurement to ensure a comparable degradation on visual-audio signal blurring (i.e. pixel distance in visual blurring and temporal alignment distance in audio blurring) [10]. While in the case of crossmodal correspondences, objective degradation in each modality does not ensure a comparable attentional capture of the same order of magnitude when modalities are in combination (i.e. perceptual sensitivity on one modality can be enhanced or inhibited by introducing another modality channel, with which combined either congruently or incongruently [54]). In this regard, there are few systematic gradation methods for balancing the correspondences across modalities. Therefore, we followed a common practice in the field of HCI, to balance the perceptual strength across modalities in a subjective manner. Four volunteered participants (age 25 - 35, two female and two male) were presented with crossmodal stimuli. The pitch, magnitude of vibration and intensity of brightness in those stimuli were varied as a function of the distance information. Then participants were asked to rate from low to high in terms of perceptual sensitivity and discernibility of graded values. We chose the parameters that have the same ratings for the experiment. Details of the choosing parameters will be explained in the next section.

Based on previous behavioural and psychophysical research, the visual, audio and vibrotactile modes were employed in different crossmodal combinations. In the audio mode, the spatial scale was mapped through the parameters of pitch [12,60] (frequency ranged from 60 Hz to 1500 Hz), in the visual mode, brightness was chosen to pair with auditory pitch [36,22] (luminous intensity ranged from [0, 0, 0] (mcd) to [800, 4000, 900] (mcd) in RGB value), and in vibrotactile mode, we used vibration intensity paired with audio frequency [3]. The intensity of the vibration is controlled by PWM (pulse width modulation) duty cycles from %0 to %100, with the amplitude ranging from 0.6 to 1.3 (g) and the frequency ranging from 120 to 260 (Hz).

We used an interactive Table-top to mark the interactive spatial boundary, and designed a manipulable tangible object displaying dynamic, concurrent crossmodal distance information. The Table-top was designed to display the graphical interface of the application, supporting the tangible object and detecting its movement, and presenting the audio when the object is active. It was built based on the reacTIVision framework, which contains one projector running at 1280×720 resolution, one Infra-red camera with a frame rate of 30Hz, two speakers placed on each side of the table below the top screen, and one laptop computer with an XBee coordinator. The application was written in a processing sketch and projected on the top screen of the table (Figure 1). Since the experimental setup required a dark environment, the table was illuminated from inside by 12 Infra-red LEDs which were integrated into the camera.

We used a teensy 3.1 as the microcontroller of the tangible object, which was embedded with one XBee router for wireless communication, one RGB LED to control the brightness and one vibration motor for vibration intensity control. We encapsulated the components into a Half-transparent plastic white box sized 5×5×5 (cm). The object was powered by a 2000 mAh rechargeable battery. We attached a fiducial marker at the bottom of the object for recognition and motion tracking.

## 4 Method

### 4.1 Experimental Design

The values of the independent variable were displaying strategies, i.e. the different crossmodal combinations employed for displaying distance to participants in a inter-subjects experimental design. Participants were allocated to one of the three groups: the unimodal group, the bimodal group and the trimodal group.

The different display conditions under each group were also considered. The three conditions under the unimodal group were the Visual-display condition (V), Audio-display condition (A), and the Haptic-display condition (H); the three conditions under the bimodal group were the Visual-haptic condition (VH), Visualaudio condition (VA), and the Audio-haptic condition (AH). The trimodal group employed a Visual-audiohaptic display strategy (VAH) (Table 1).

**Table 1** Experimental conditions.

| Groups | Conditions | | |
|---|---|---|---|
| Unimodal | V (visual) | A (audio) | H (haptic) |
| Bimodal | VA (visualaudio) | VH (visualhaptic) | AH (audiohaptic) |
| Trimodal | VAH (visual-audio-haptic) | | |

4.2 Experimental Task

Participants were instructed to search for the invisible target hidden in the periphery of the table surface. They were told to focus on movement accuracy (i.e. avoid random movement during the search) rather than speed of completion during the search task. The initial layout of the application was a 9 × 9 grid as shown in figure 1 (right). Before each trial began, participants placed the physical object in the red square (Figure 1), the centre of the grid, to start a new trial. When participants moved the object around the table surface searching for the target, feedback about the current Euclidean distance of the object from the target was displayed through either unimodal or crossmodal feedback. Based on perceived distance, participants estimated target location and adjusted arm movements accordingly. Once participants move the object into the square where the target is hidden, the system presents an auditory icon

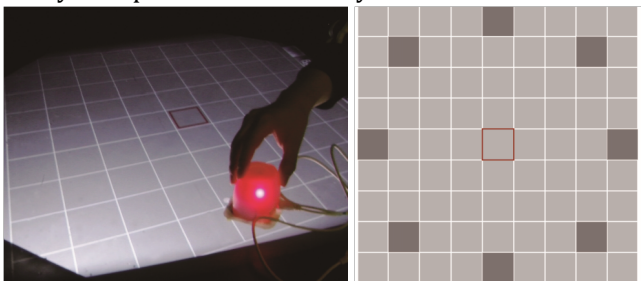

**Fig. 1** The workbench (left) and the Table-top interface (right). The dark grey area represents the potential positions of the hidden target, which were not visible during the experiment.

(of 1s duration) to inform the participant of success and that the trial is over.

In order to reduce the learning effects between trials, the target was allocated to random positions at each trial with no repetition, while the distance between the target and the starting point (the centre of the grid) always remained the same (Figure 1 (right)). Each participant in all three groups performed eight trials under the same condition with the same experiment level. There was no baseline condition without feedback, because the task would become difficult without any feedback, since participants would only be able to guess randomly at the location of the target. We do not believe this would provide a meaningful baseline for the task.

4.3 Participants

Thirty-four participants were recruited to take part in the study. Four outliers were excluded. Two of the participants could not distinguish the brightness levels due to vision weakness and short sight, and another two produced very erratic movements and could not complete the task in a reasonable time. Finally, thirty participants (15 male, 15 female) were included in the latter quantitative analysis. The participants' ages ranged from 24 to 45 years old (M = 29.1, SD = 5.5). They were recruited through university mailing lists as well as personal contact lists. They were a mixture of nationalities, studying a variety of degree courses or in different working fields. All participants volunteered to take part and received a snack for their involvement.

4.4 Procedure

First, participants were given a consent form with details of the experiment to read and sign. After that, they were asked to complete a questionnaire, including demographic information, music training history, and experience of interactive table applications. The results of the questionnaire showed that fourteen of the thirty participants had music training. Seventeen of the thirty participants had experience of interactive table applications. Participants were then assigned to each group to achieve an approximate balance based on the above information and their working field. We instructed participants to find the hidden target as accurately as possible by moving the object around the table. We informed participants that the target position was randomised, and that feedback information would change in real-time based on their distance to the target, but we did not explain the crossmodal correspondences under exploration.

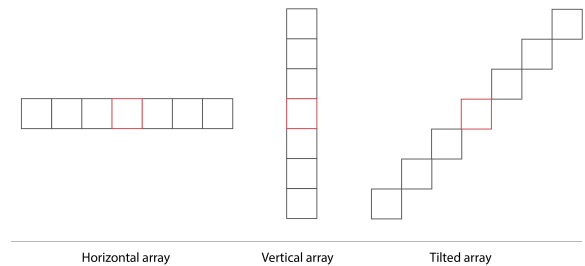

**Fig. 2** The graphic layouts in the warm-up application.

Before the experimental trials began, participants used a warm-up application employing the same apparatus. However, instead of using the 2-dimensional grid that was used in the experimental trials, participants were trained in a 1-dimensional layout including a horizontal array, vertical array, and a 45 degree tilted array (Figure 2). In order for them to be familiar with the range of the modal variables but without overtraining, only the unimodal feedback mode was used in each of the three layouts respectively.

The conditions for the unimodal groups and bimodal groups were counterbalanced using a 3×3 Latin square design. Participants were permitted to pause at any time between each trial. Finally, we administered a short questionnaire collecting subjective ratings concerning the interaction and any free form comments. The average duration of each experiment was 32 mins.

4.5 Measurements

With the aim of addressing the research questions, two types of measurements were collected respectively. The first question is 'whether displaying distance information through crossmodal feedback in target-searching task support better performance than that through unimodal feedback?'. Usability evaluation was used for addressing this question, which includes both quantitative and qualitative measurements. The quantitative measurements taken were task efficiency and accuracy. Few searching steps indicate good efficiency. A searching step was counted every time that a participant moved the object from one grid square to another. Fewer erroneous steps indicate better accuracy. An erroneous move was designated whenever a participant moved the object in such a way that either the X or Y coordinate of the object went in the opposite direction to the location of the target. If both X and Y coordinates were erroneous, this counted as just one erroneous move. Both the searching and erroneous steps were collected through the system's log files. The navigation time of each trial was also collected. The qualitative measurement was subjective evaluation collected through postexperimental questionnaires.

The second question this paper tried to address is 'how the strategy used to display distance influences the target-searching performance'. To address this question, usability evaluation is not sufficient since it mainly reflects interaction results rather than interaction process. To understand what was happening within the interaction, a kinematic analysis was conducted to unveil movement features [4,51,49]. Firstly, the searching trajectories were quantified by vectorising searching trajectories (figure 3 (left)), which was calculated based on the absolute angle of each vector. Secondly, we measured the steering angles of searching movements (figure 3 (right) and 4), which was calculated based on the relative angle between two consecutive moves. By comparing the vector and steering angle distributions between different feedback conditions, we were able to observe how different strategies for displaying distance modulated target-searching behaviour. Thirdly, the length of searching trajectories was collected as an indicator of searching effort. The shorter the distance, the less effort the participants needed to expend in searching for the target.

4.6 Hypotheses

The following three hypotheses were formed to address the first question:

H 1: Participants in the bimodal and trimodal groups will reach the target with fewer searching steps than those in the unimodal group.

H 2: Participants in the bimodal and trimodal groups will reach the target with fewer erroneous steps than those in the unimodal group.

The following hypothesis was formed to tackle the second question:

H 3: The movement features made during the tasks within the unimodal and crossmodal groups will appear

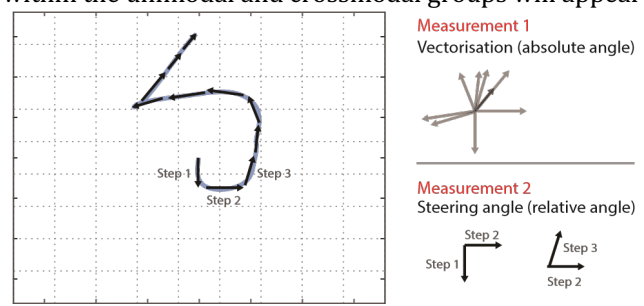

**Fig. 3** Quantification of the trajectory (left). Measurement 1 (top-right) is the density of the gesture vectors, which was calculated according to the absolute angle of each vector. Measurement 2 (bottom-right) is the gesture steering angle, which was calculated according to the relative angle between two consecutive movements.

**Fig. 4** The steering angles represent four types of gestural movement: make forward, make diversion, make turn, and reverse.

differently in terms of the gesture vector and steering angle distributions.

## 5 Results

### 5.1 Usability Analysis

A Kolmogorov-Smirnov test was run, which showed that the data of searching steps and erroneous steps could be assumed to be normally distributed. We ran One-way ANOVAs to compare the numbers of searching steps and numbers of error steps under the unimodal, bimodal and trimodal display conditions. The Fisher's LSD test was used for Post hoc tests of main effects. We used a confidence level of $\alpha = 0.05$ for the tests.

*5.1.1 Searching efficiency: searching steps across groups*

*Conditions involving auditory display*

There was no significant main effect on searching steps between the unimodal (Audio-mapping) condition, the bimodal (Audio-haptic) condition and the trimodal (Audio-haptic-visual) condition ($F_{2,237} = 0.07, p =$ haptic) condition and the trimodal condition, we also found a main effect ($F_{2,237} = 5.43, p = .01$). Post hoc tests showed that participants in the bimodal condition (*mean = 12.76, sd = 6.53*) spent significantly fewer steps than participants in the unimodal condition (*mean = 15.16, sd = 5.89*)(*p = .012*). Participants in the trimodal condition (*mean = 12.24, sd = 5.51*) also spent fewer steps than those in the unimodal condition (*mean = 15.16, sd = 5.89*)(*p = .002*). These results show that in the conditions involving vision, both the bimodal and trimodal mappings increased the efficiency of interaction.

*Conditions involving vibrotactile display*

There was a significant main effect on searching steps between the unimodal (Haptic-mapping) condition, the bimodal (Haptic-audio) condition and the trimodal condition ($F_{2,237} = 7.18, p = .00$). Post hoc tests showed that participants in the bimodal condition (*mean = 12.11, sd = 4.77*) spent significantly fewer steps than participants in the unimodal condition (*mean = 15.99, sd = 5.93*)(*p = .001*). Participants in the trimodal condition (*mean = 12.24, sd = 5.51*) also spent fewer steps than those in the unimodal condition (*mean = 15.99, sd = 5.93*)(*p = .00*).

There was also a significant effect on searching steps between the unimodal condition, the bimodal(Hapticvisual) condition and the trimodal condition ($F_{2,237} = 4.73, p = .01$). Post hoc tests showed that participants in the bimodal condition (*mean = 12.76, sd = 6.53*) spent significantly fewer steps than participants in the unimodal condition (*mean = 15.99, sd = 5.93*)(*p = .02*). Participants in the trimodal condition (*mean = 12.24, sd = 5.51*) also spent fewer steps than those in the unimodal condition (*mean = 15.99, sd = 5.93*)(*p = .00*).

*Conditions involving auditory display*

We found a main effect on erroneous steps between the unimodal (Audio-mapping) condition, bimodal (Audiovisual) condition and trimodal condition ($F_{2,237} = 3.46, p = .03$). Post hoc tests showed that participants in the bimodal mapping (*mean = 4.35, sd = 4.38*) condition made fewer erroneous steps than they did in the trimodal mapping condition (*mean = 6.26, sd = 5.09*).

*Conditions involving visual display*

There was a main effect of erroneous steps between the unimodal (Visual-mapping) condition, bimodal (Audiovisual) condition and trimodal condition ($F_{2,237} = 6.77, p = .00$). Post hoc tests showed that participants in the bimodal mapping condition (*mean = 4.35, sd = 4.38*) made fewer error steps than they did in the unimodal mapping condition (*mean = 6.87, sd = 4.05*) and trimodal mapping condition (*mean = 6.26, sd = 4.64*).

*Conditions involving vibrotactile display*

There was no significant main effect between the unimodal (Haptic-mapping) condition, bimodal (Hapticvisual) condition and trimodal condition ($F_{2,237} = 0.37, p = .69$), and there was no significant main effect between the unimodal condition, bimodal (Haptic-audio) condition and trimodal condition ($F_{2,237} = 0.13, p = .88$).

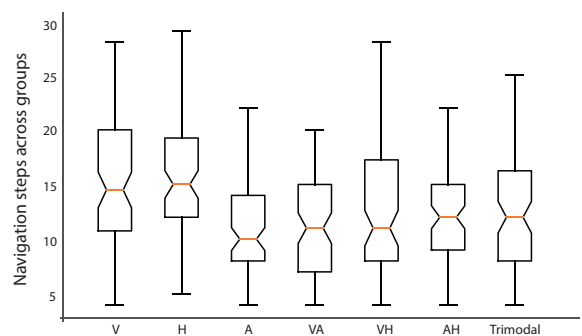

**Fig. 5** Searching steps in the unimodal and crossmodal conditions. Error bars represent standard deviation.

*5.1.3 Navigation time analysis*

Individual difference in movement style in this targetsearching task was large. Some participants performed very fast with smooth arm motions, while others performed with slow and jerky movements with intermittent pauses. Given this fact, statistical analysis on the navigation time would not lead to meaningful insights. While considering there might be a time-error trade-off, we calculated the average navigation time of each of the

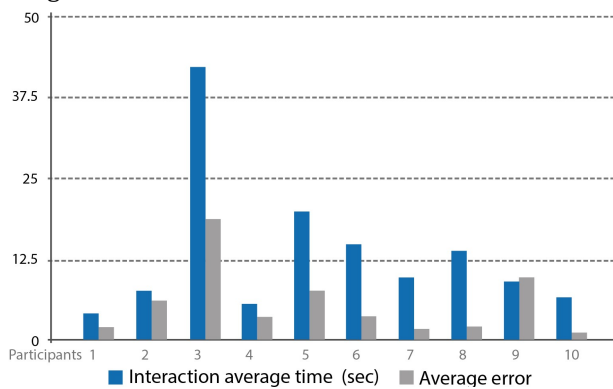

**Fig. 6** and error comparison in the visual-haptic mapping condition.

thirty participants. The results showed no evidence of a general time-error trade-off in the data, though occasional specific trials reflected this trade-off. Figure 6 presents one of the examples under the visual-haptic mapping condition. The instructions given to participants at the beginning of the experiment, which ask them to focus on task accuracy, might account for this fact. While if participants were instructed to focus on the time instead of accuracy, the time-error trade-off might be observed.

5.2 Kinematic Analysis

Kolmogorov-Smirnov test showed that the three kinematic measurement data-sets were not normally distributed, thus non-parametric statistical techniques were applied. To quantify movement features without confounding of different target positions, all trials have been rotated towards the same direction, the north, for analysis.

*5.2.1 Motion feature 1: Gesture vector distribution*

The first measurement of this kinematic analysis was gesture vector distribution. Gesture trajectories can be observed directly in the plot figure 7, from which the differences between the unimodal conditions and the crossmodal conditions can be intuitively identified. The gestures in the unimodal conditions and the trimodal condition tend to be made with random movements without convergent trajectories or clear movement pattern, while the gestures in the bimodal conditions showed a salient directional tendency, which converged in the orthogonal direction (after rotation). The same movement feature can be observed in other plots as well.

The gesture vectors quantify the above observation. By observing the vector distribution in different angles, gesture trajectories can be easily evaluated in terms of the frequency in angle distributions. For example, in figure 8, the vector distribution in the visual condition has a smaller ratio of orthogonally aligned vectors than that in either the visual-auditory condition or the visual-haptic condition. This evaluation from a density plot corresponds to the original gesture trajectories in other conditions as well.

Figure 8 shows examples of three distributions under the A, VA and VAH conditions. The percentage of the vectors in eight directions is listed in table 2. In the unimodal conditions, the frequency of vector distribution along orthogonal directions for the V, A and H conditions was 60.41 %, 58.82%, and 59.80 %. In the bimodal conditions, the frequency for VA, VH and AH conditions were 77.30%, 75.68%, and 70.96%. In the trimodal condition, the frequency was similar to that in the unimodal conditions, which was 61.67%.

For the vector distribution along the diagonal directions, the unimodal conditions, V (39.59%), A (41.18%) and H (40.204%) conditions have higher frequency than that in the bimodal conditions, the VA (22.70%), VH (24.32%), and AH (29.04%). The frequency in the trimodal condition was similar to that in the unimodal conditions (38.33%).

These results quantitatively represent the observed difference between the display conditions as shown in figure 7. However, a Kruskal-Wallis test showed that there were no significant difference between the unimodal, bimodal and trimodal feedback conditions ($H(2) = .42, p = .81$).

*5.2.2 Motion feature 2: Steering angle distribution*

The second measurement of the kinematic analysis was the frequency of steering angles. Figure 9 shows the examples of steering angle distribution under the A, VA, VHA conditions. The full results under seven conditions are listed in the table 3. The three unimodal conditions, the visual, auditory and haptic conditions, have relatively higher distributions between bin 100° to 170°, which indicates the gesture of making turns (figure 4), and lower distribution between bin 170° to 180°, which indicates reversed gestures movements, than the three bimodal conditions (the VA, VH and AH conditions). The three bimodal conditions have fewer turns and more reversed movements than the three unimodal conditions. The distribution of the trimodal condition falls between the unimodal conditions and the bimodal conditions.

A Kruskal-Wallis test was applied to compare the difference in steering angle distributions between the unimodal, bimodal and trimodal groups. The results showed that steering angle distributions were significantly affected by display strategies ($H(2) = 69.62, p = .00$). A Bonferroni correction was applied and so all effects are reported at the 0.167 level of significance. Post hoc test showed that the steering angle distribution under the bimodal feedback conditions was significantly different to that under the unimodal conditions ($U = 68.38, r = -0.16$), as well as that under the trimodal condition ($U = 17.08, r = -0.10$). However, there was no difference between the unimodal conditions and trimodal condition ($U = 2.85, r = -0.05$).

*5.2.3 Motion feature 3: Searching trajectory lengths*

The trajectory length reflects how much effort participants made during the searching task. Results showed that participants made the shortest trajectories under the VA display condition, followed with the A, AH, VH, VHA, V and H conditions (figure 10).

A Kruskal-Wallis test showed a significant difference in the trajectory lengths between the unimodal, bimodal and trimodal display conditions ($H(2) = 7.09, p = .029$). A Bonferroni correction was applied, and so all effects are reported at the 0.167 level of significance. Post hoc test showed that the trajectory lengths under the bimodal conditions were shorter than the lengths produced under the unimodal conditions ($U = 5341.00, r = −0.30$).

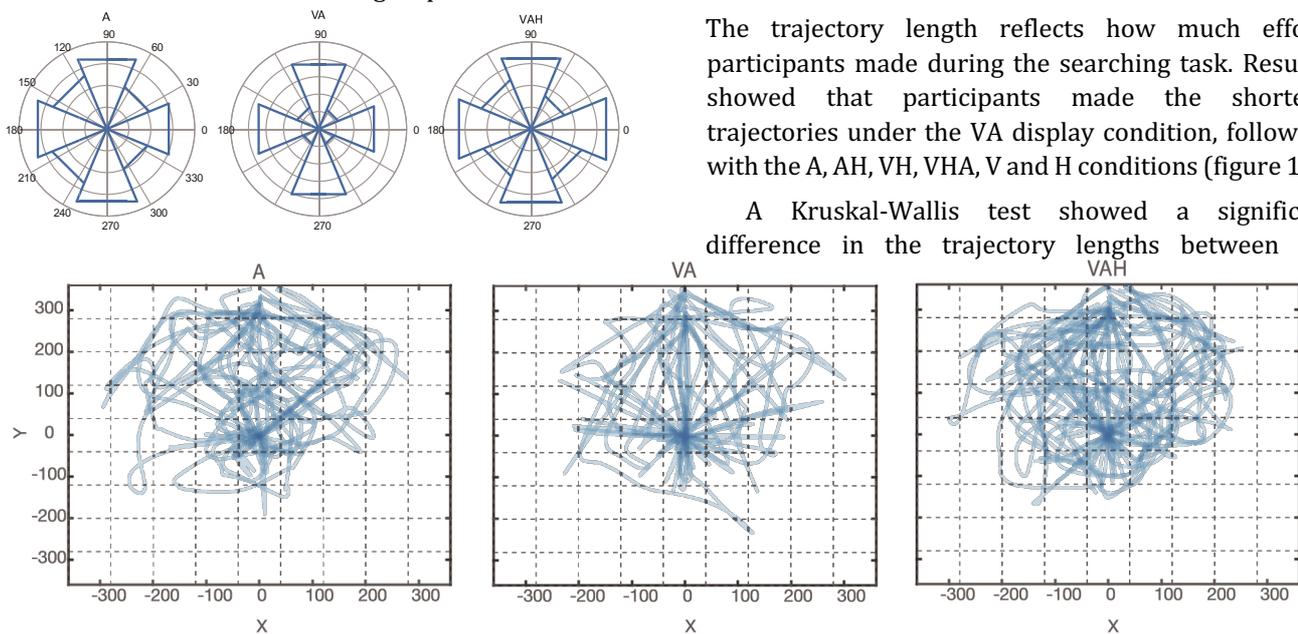

**Fig. 7** Examples of searching trajectories under the audio display condition (left), visual-audio condition (middle) and trimodal condition (right). Coordinate unit is in millimeter.

**Table 2** Motion vector distributions

| Direction | V(%) | A(%) | H(%) | VA(%) | VH(%) | AH(%) | VAH(%) |
|---|---|---|---|---|---|---|---|
| N | 14.26 | 15.03 | 12.95 | 20.41 | 16.14 | 18.56 | 14.99 |
| W | 16.70 | 14.82 | 16.18 | 18.88 | 21.36 | 16.81 | 15.20 |
| S | 14.63 | 15.69 | 14.31 | 20.66 | 16.36 | 20.09 | 15.63 |
| E | 14.82 | 13.29 | 16.35 | 17.35 | 21.82 | 15.50 | 15.85 |
| Orthogonal | 60.41 | 58.82 | 59.80 | 77.30 | 75.68 | 70.96 | 61.67 |
| NE | 7.50 | 9.15 | 8.86 | 6.12 | 5.23 | 5.46 | 7.07 |
| NW | 9.57 | 11.55 | 9.71 | 6.89 | 4.32 | 9.83 | 10.49 |
| SW | 10.51 | 11.77 | 9.71 | 4.85 | 8.64 | 5.24 | 10.06 |
| SE | 12.01 | 8.71 | 11.93 | 4.85 | 6.14 | 8.52 | 10.71 |
| Diagonal | 39.59 | 41.18 | 40.20 | 22.70 | 24.32 | 29.04 | 38.33 |

**Fig. 8** Rose plot of gesture vector (absolute angle) distribution under seven conditions. The right figure shows trimodal condition, the middle figure shows VA bimodal condition, and the left line shows A unimodal condition.

Given the observation that the conditions with audio display have similar trajectory lengths across uni, bi and trimodal conditions, we run the Kruskal-Wallis test to make comparisons between A, VA, AH and VAH conditions. Result showed that there was no statistically significant difference in trajectory lengths between those conditions ($H(3) = 1.92, p = .59$). In summary, these results reflect that the bimodal conditions elicited

## 6 Discussion of Usability Evaluation

Hypothesis 1 that participants in the crossmodal groups will reach the target with fewer searching steps than those in the unimodal group has been confirmed in the conditions involving visual and haptic displays and rejected in the conditions involving auditory display.

Hypothesis 2 that participants in the crossmodal groups will reach the target with fewer erroneous steps than those in the unimodal groups has been confirmed only in the comparison between visual-only and visualaudio display conditions. However, the bimodal group produced significantly fewer erroneous steps than the trimodal group did with the audio and visual display.

These results reflect that crossmodal display conditions elicited more efficient target-searching without reduced erroneous steps. Considering there was no concurrent feedback of target direction at the beginning of each trial, it is expected that we would observe similar explorative movement patterns across conditions at the start of the trial. Thus all participants had an equal possibility to make erroneous shortest trajectories in general. However, in terms of trajectory length, audio display elicited the same good performance as that in crossmodal conditions.

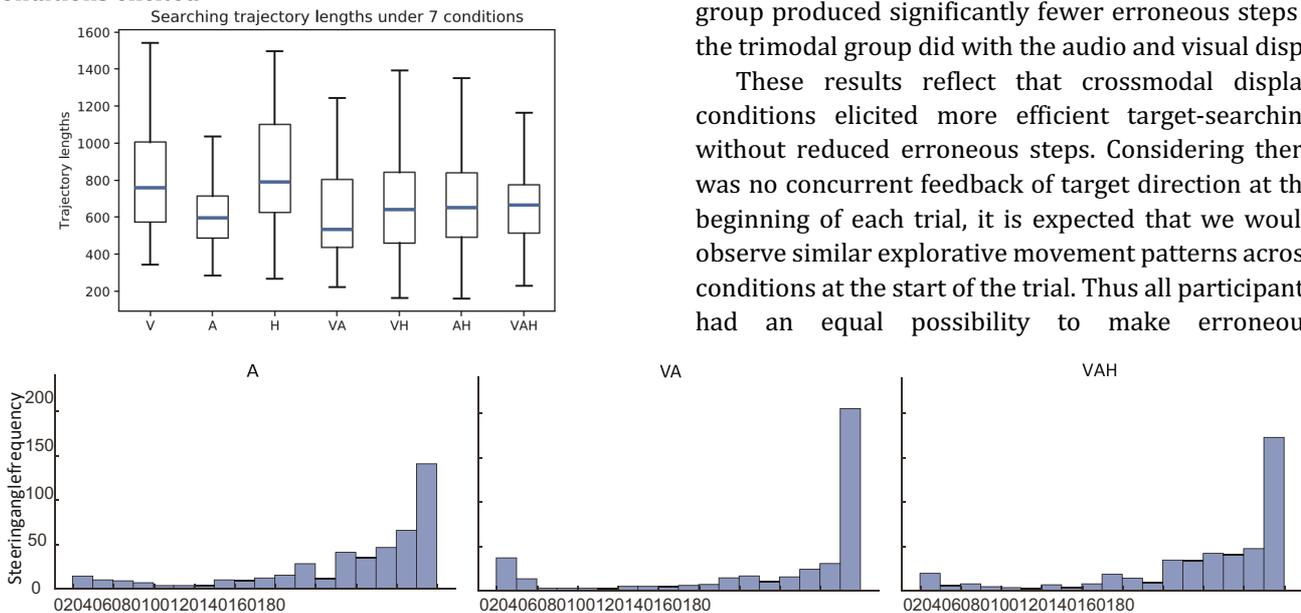

**Fig. 9** Steering angle (relative angle) distribution under the A, VA and VAH conditions.

**Table 3** Steering angle distributions

| Steering angles | V(%) | A(%) | H(%) | VA(%) | VH(%) | AH(%) | VAH(%) |
|---|---|---|---|---|---|---|---|
| $pi * 0$ | 2.74 | 4.18 | 1.24 | 4.61 | 3.97 | 3.64 | 2.93 |
| $pi * (1/10)$ | 2.35 | 2.09 | 1.42 | 0.81 | 1.64 | 2.27 | 1.81 |
| $pi * (2/10)$ | 2.35 | 0.93 | 2.66 | 1.08 | 1.40 | 1.59 | 1.35 |
| $pi * (3/10)$ | 2.35 | 2.78 | 1.95 | 2.17 | 1.64 | 1.59 | 2.26 |
| $pi * (4/10)$ | 2.94 | 4.64 | 4.96 | 2.17 | 3.05 | 2.50 | 4.29 |
| $pi * (5/10)$ | 8.02 | 6.73 | 9.57 | 3.79 | 5.37 | 2.95 | 5.19 |
| $pi * (6/10)$ | 12.13 | 9.51 | 11.17 | 6.50 | 6.07 | 6.59 | 11.51 |
| $pi * (7/10)$ | 16.63 | 15.31 | 15.60 | 6.78 | 8.18 | 8.18 | 14.90 |
| $pi * (8/10)$ | 21.14 | 22.51 | 20.75 | 13.55 | 12.38 | 18.18 | 17.38 |
| $pi * (9/10)$ | 25.83 | 28.77 | 28.90 | 50.41 | 51.40 | 45.00 | 34.76 |
| $pi$ | 3.52 | 2.55 | 1.77 | 8.13 | 4.44 | 7.50 | 3.61 |

**Fig. 10** Trajectory lengths produced under seven conditions.

movements. This fact indicates two things: firstly, it is reasonable to say that participants in the crossmodal display required fewer correction steps to get back in the direction of the target, which means that participants in the crossmodal group expended less effort in discovering they had deviated from the direction of the target. Secondly, since the crossmodal display did not reduce the number of erroneous steps, it is fair to say the concurrent crossmodal display could not facilitate people keeping in the correct direction with scalar information alone.

In spite of the crossmodal advantages for task efficiency, the audio display in the unimodal group showed a statistically equivalent good performance to crossmodal display conditions. A previous research used discrete corssmodal feedback revealed a similar behavioural pattern, which was due to the auditory dominance effect. Though in our current study, we used continuous crossmodal display [41], it is plausible to deduce that the auditory dominance effect may play a part as well. The auditory dominance effect could account for this observation [48], in that auditory perception of the changing pitch is the most facilitatory cue that compensated other modality channels [20]. While auditory feedback may provide more robust and unambiguous distance information than visual or tactile feedback does, there is no evidence showing that this is due to a finer pitch mapping. On the one hand, as the results show in section 5.1.1, without the pitch mapping, the VH condition produced statistically equally good performance with the VHA condition. On the other hand, subjective reflections collected from the Post-experiment questionnaire provide insight on pitch mapping. The ratings of the modality the participants relied on most in the crossmodal groups are shown in figure 11. According to their responses, the continuous changing pitch was good for distance estimation and helped them to remember the route that they had already searched. The objective result, as well as the subjective reflection, are consistent with previous research, in which the pitch was used for distance mapping in a topography exploration task [45], and in a similar target searching task [38]. Although there was no difficulty in following the crossmodal displays for both the brightness and the vibrotactile dimensions, the answers to the questionnaire show that the real-time feedback from those two modes was not ideal for facilitating spatial memory during exploration, and lead to more random estimations of the distance.

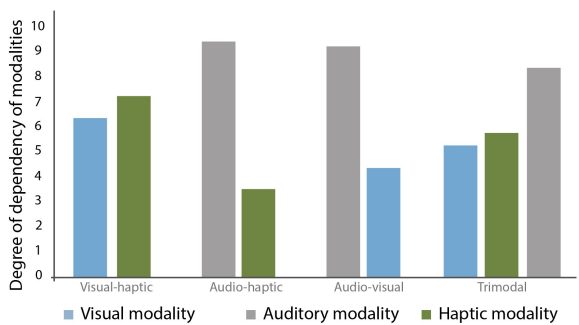

**Fig. 11** Subjective rating on modality preference in crossmodal conditions. Rating scale ranging from 1 to 10.

In the trimodal condition, two people reported that they sometimes ignored the haptic (vibrotactile) display intentionally because they found it distracting, and one person reported that he ignored the visual brightness display the whole time just for concentration. The statistical results show that most trials in the trimodal condition yielded similar good performance as that in the bimodal condition, but it elicited more erroneous steps. One interpretation of this might be that too much concurrent information display in relation to a single task might cause cognitive overload and compromise the crossmodal benefit for task efficiency.

In summary, the usability evaluation showed that the bimodal display conditions supported the best task efficiency and good accuracy in general, while the audio display compensated the unimodal disadvantage and made the audio display condition equally good in the crossmodal conditions (i.e. VA, AH, VAH). The trimodal condition elicited a similar level of efficiency as the bimodal conditions did but produced more erroneous moves due to cognitive overload. The subjective reports of participants concerning their reliance on the audio and trimodal interaction experience were consistent with the behavioural data observed. However, these subjective reports cannot explain why the crossmodal conditions and the conditions with audio display improved searching efficiency. In this regard, we evaluated motion features based on the kinematic analysis, which is discussed below.

**7 Discussion of Kinematic Evaluation**

The hypothesis H3 that the movement features produced under the unimodal conditions and the crossmodal conditions will appear differently was partially confirmed. The steering angle distribution had a statistically significant difference between the bimodal conditions and the unimodal conditions, as well as the trimodal condition. However, the trimodal condition was not statistically different from the unimodal conditions. These results indicate that the crossmodal display, with the feedback in the form of visual brightness, auditory pitch and vibrotactile intensity in particular, spontaneously modulates the arm movements in a way that produce more efficient searching performance.

The analysis based on the steering angle distribution explained the results, at the kinematic level, that bimodal conditions produced the best task efficiency and good accuracy. Specifically, the steering angle at 0° and 180° in bimodal conditions have a higher frequency than that in the unimodal conditions as well as trimodal condition (figure 9). These two angles corresponded to the actions of moving forward and moving in reverse respectively as shown in figure 4. The result of the usability evaluation showed that the bimodal display conditions facilitated the correction of movements rather than guiding in the right direction in the first place. Given this observation, the gesture motion of 'move forward' and 'move in the reverse' direction were more likely to have occurred as corrective movements. As a result, the increased number of these two types of gestural movements turned out to be a better strategy for searching and approaching the target than making turns or diversion. The third kinematic measurement, the searching trajectory length, reflected that the crossmodal group produced the shortest length compared with the trimodal and unimodal groups, not the audiodisplay condition.

It is reasonable to deduce that the improved task efficiency and reduced effort in bimodal conditions, in general, was due to an improved movement strategy that was supported by the bimodal display strategy. Given the fact that participants could not see their own trajectories during the task, conscious adjustment of their arm movements according to the feedback is unlikely to happen. However, whether participants were truly unaware of their behavioural change needs to be confirmed with further investigation.

The gesture vectors distribution, e.g. the quantification of the gestures, however, did not show a statistical significance across unimodal, bimodal and trimodal groups, though the gesture plots clearly showed a different tendency between movements, either more a focused direction or randomly (figure 7). One possible explanation might be that due to the nature of the nonparametric statistical test, which compared the ranked data rather than the original data, some motion features may have been lost during the process of ranking. This disadvantage could theoretically be compensated to some extent by increasing sample size. The current experiment may not have the necessary sample size to be large enough to discover a significant effect between the feedback strategies. Another possible reason might be that the method of quantifying those gestures may not be accurate enough to reflect the differences. To improve the acuity, time-series data with higher resolutions need to be collected for further investigation.

## 8 General Discussion

This paper investigated concurrent unimodal and crossmodal display of distance information in the targetsearching task. The usability and kinematic evaluations addressed the research questions whether and how concurrent crossmodal display through a tangible interface influence people's target-searching performance. The usability evaluation results showed that the bimodal display supports the best searching efficiency and accuracy compared with the unimodal and trimodal display. However, the audio display condition (unimodal condition) supported a statistically equally good searching performance as the bimodal display conditions. The kinematic evaluation confirmed the usability evaluation in terms of the searching trajectory length. Meanwhile, the steering angle measurement explained the usability results in that the bimodal display conditions induced a more efficient movement strategy, which includes a high frequency of 'moving forward' and 'make reverse' movements. This strategy reduced the searching steps and trajectory lengths, i.e. improved task efficiency with reduced searching effort.

### 8.1 Contributions and Design Implications

There are two main contributions of this paper. Firstly, to our knowledge, this paper provides the first empirical evaluation of concurrent visual, vibrotactile and auditory, as well as combined display strategies, which was instantiated through a tangible interface designed for the target-searching task. Secondly, this research contributed empirical evidence on concurrent crossmodal display, and employed kinematic analysis to understand gesture-based movements in a target searching task. Considering the results altogether, we summarise the following implications that could inform future interface design and research studies.

– Assisting grasping and maximizing accuracy

Usability evaluation showed that the VA display induced the best performance in terms of target-searching efficiency, and was followed by audio, AH, VH, VHA, visual and haptic display. The kinematic analysis also showed that the display conditions with audio display did not have a statistical difference in searching efficiency and efforts. These results imply that the audio display for distance information can support comparable good performance even in the absence of visual (brightness) display. Haptic (vibration intensity) display alone, in comparison, do not have such a facilitatory effect, but when combined with visual display, the crossmodal effect leads to improved searching efficiency and effort than that of both the visual and haptic unimodal displays.

Based on these results, we propose that the auditory display could thus be used for helping visually-impaired people explore the items on a table or a workbench, where the exact location or direction of the items is uncertain. Although there are existing applications that were designed with a similar idea [6,16], the results in these previous studies were collected on a desktopbased prototype with mouse cursor movements [6,7], or with a different interaction intention [44,16]. Our experiment made a step closer to the real-world scenario, analysed target-searching behaviour on the table, which provide empirical evidence to inform the design of assistive technology.

In other cases, the crossmodal or auditory display could also be applied to maximise the motor accuracy in the scenario of motor learning, remote control or surgical operation. In a similar line of thinking, multi-modal feedback has been proposed as an effective approach for data exploration tasks [24], and has been shown to facilitate learning new trajectories [9], locate objects as well as improve spatial precision [10]. Based on these empirical accumulations, our results further suggests that crossmodal correspondences have an effect on modulating searching movements, which in turn improves searching efficiency and reduces trajectory lengths. While in these applications, the number of display channels, feedback intensity as well as people's motor ability and awareness needs to be taken into account in a specific interaction context.

– Decoding interactive motion features

In this paper, we used kinematic analysis to understand the difference in target searching movements under different crossmodal display conditions. This analysis confirmed the usability evaluation results, in addition, explained why and how the searching performance varies in terms of kinematic features. Thus we propose that kinematic analysis, with the focus on evaluating motion quality, can be a complimentary evaluation technique for understanding interactive behaviour, particularly where the tasks require careful manipulation and consecutive or continuous interactive movements [47,27,19].

– The proper number of displaying channels

Although crossmodal correspondence is rooted in our multimodal perception of real-world phenomena, results obtained from both the usability evaluation and the kinematic analysis showed that crossmodal display using all three modalities might not produce the best interaction performance in terms of searching efficiency, accuracy and effort. Although too much instantaneous feedback through one tangible object may be cognitively demanding and may lead to more errors than when fewer modes are used, this result points to another investigation possibility regarding the embodied physicality of tangible interfaces. For the future investigation, we suggest reducing the cognitive load by layering discrete feedback on continuous crossmodal display for target searching tasks.

8.2 Limitations and Future Work

The first limitation, as discussed in section 7, is that the vectorisation of original movement trajectories did not reflect the observable difference between display strategies. Future studies could either increase the sample size or improve the resolution of data collection to verify the validity of the evaluation method.

Secondly, the current research did not include the display intensity as a manipulation factor. We intentionally did not manipulate the feedback intensity to avoid potential confounds and interaction effects between the feedback groups and display intensity. As such, the current research outcomes cannot apply to situations where the relative intensity between modalities is continuously changing. As the relative intensity is also an important feature that influences information processing, future research should further investigate this factor with the same experimental approaches.

Last but not least, the results of the current study was obtained based on scalar information (i.e. distance). To increase the feasibility of crossmodal feedback in a more ecological target-searching task, directional/vector information should be included as well. Build on the current stage of the investigation, we evolved the current system which integrated the distance and directional information. The near future investigation will focus on optimised crossmodal combination strategy in the context of target-searching for the design of assistive technology.